Original Research Paper

# Conceptual Modeling for Computer Organization and Architecture

**Sabah Al-Fedaghi**

*Department of Computer Engineering, Kuwait University, Kuwait City, Kuwait*

**Abstract:** Understanding computer system hardware, including how computers operate, is essential for undergraduate students in computer engineering and science. Literature shows students learning computer organization and assembly language often find fundamental concepts difficult to comprehend within the topic materials. Tools have been introduced to improve students' comprehension of the interaction between computer architecture, assembly language, and the operating system. One such tool is the Little Man Computer (LMC) model that operates in a way similar to a computer but that is easier to understand. Even though LMC does not have modern CPUs with multiple cores nor executes multiple instructions, it nevertheless shows the basic principles of the von Neumann architecture. LMC aims to introduce students to such concepts as code and instruction sets. In this paper, LMC is used for an additional purpose: a tool with which to experiment using a new modeling language (i.e., a thinging machine; TM) in the area of computer organization and architecture without involving complexity in the subject. That is, the simplicity of LMC facilitates the application of TM without going deep into computer organization/architecture materials. Accordingly, the paper (a) provides a new way for using the LMC model for whatever purpose (e.g., education) and (b) demonstrates that TM can be used to build an abstract level of description in the organization/architect field. The resultant schematics from the TM model of LMC offer an initial case study that supports our thesis that TM is a viable method for hardware/software-independent descriptions in the computer organization and architect field of study.

## Introduction

Understanding computer system organization and architecture is essential for undergraduate students in computer engineering and science (Kurniawan and Ichsan, 2017; Ledin, 2020). Computer organization denotes the level of abstraction above the level of digital logic, but below the operating system level (Berger, 2012). At this level, the main components are functional units that correspond to particular constituents of hardware (e.g., memory, input, output, arithmetic-logic unit, and control unit) built from lower-level blocks. Different levels of abstraction can also be applied to computer architecture. On the software level, computer architecture refers to the assembly language systems that connect the various parts of a computer's hardware into a single functioning system. "When dealing with hardware, it applies equally to the methods of creating and utilizing hardware and the process of constructing computer components" (McGee, 2020).

According to Yurcik and Brumbaugh (2001), students learning computer organization and assembly language often finds fundamental concepts difficult to comprehend (Black and Komala, 2011; Yurcik and Brumbaugh, 2001). Traditionally, students learned computer organization/architecture by studying specific hardware; however, a continuing trend in the field requires an abstract ability and the employment of more simulation tools (Decker, 1985; Bindal, 2019).

According to Wolffe et al. (2002), as real computers become more complex, they become less suitable for teaching the concepts typically found in introductory computer organization courses. In such a situation, simple hypothetical machine simulators can serve an important role by giving students access to the internal operation of a system. Hypothetical machine simulators excel at illustrating core concepts such as the von Neumann architecture, the stored program concept, and sequential execution (Wolffe et al., 2002; Nova et al., 2013).

*Little Man Computer*

Tools have been introduced to improve students' comprehension of the interaction between computer architecture, assembly language, and the operating system. One such tool is the little man computer (LMC) model of a computer that operates, roughly, in a similar way to a real computer but that is easier to understand (Englander, 2014).

LMC is that most enduring simple hypothetical machine that dates back to MIT in the 1960s (Englander, 2014; Englander, 2009; Thurgood, 2018). According to Yurcik and Osborne (2001), the LMC paradigm has stood the test of time as a conceptual device that helps students understand the basics of how a computer works. "With the success of the LMC paradigm, LMC simulators have also proliferated" (Yurcik and Osborne, 2001). Yurcik and Osborne (2001) described the evolution of five different LMC simulators from their text-based origins to the current graphical and Web-based versions. Even though LMC does not have modern CPUs with multiple cores nor execute multiple instructions, it nevertheless shows the basic principles of von Neumann architecture. It is still in use as an educational architecture for teaching the basics of assembly language. LMC is more a model or a paradigm and its aim was to introduce students to such concepts as code and instruction sets.

*Objectives*

In this paper, a high-level abstraction is introduced to model LMC that also provides a new way for using LMC for whatever purpose (e.g., education or a thinking and programming toy). We claim this LMC model is more systematic (e.g., uses only five verbs) and richer than other LMC representations developed over the last 50 years.

Additionally, this paper aims at the long-term purpose of introducing a new modeling language (i.e., thinging machine; TM) in the field of computer organization and architecture. For such an objective, LMC is used as a vehicle to show the features of TM when applied in the area of computer organization and architecture without involving complexity in the subject. That is, LMC's simplicity facilitates the application of TM without going deep into computer organization/architecture materials. Accordingly, the paper's contribution is demonstrating that TM can be used to build an abstract specification level in the organization/architect field.

The next section briefly describes the TM modeling language to achieve a self-contained paper. More details of TM can be found in Al-Fedaghi (2020a-d), but the example in the section is a new contribution. Because LMC information is available on the Internet, Section 3 introduces a summary of the important concepts related to this paper. Section 4 contains three models of LMC: static, dynamic, and behavioral. In Section 5, we explore some features of these models.

## Thinging Machine Modeling

The term *thinging* in TM comes from Heidegger (1975), in which thinging expresses how a "thing things," which he explained as gathering or tying together its constituents. *Machine* refers to the abstract machine shown in Fig. 1, which is a generalization of the known input-process-output model. The input and output are lumped together in the *transfer* stage (Fig. 1) that represents the machine's gate, where things flow to/from other machines. The *release* stage is a waiting stage for things in the machine until transferring is activated (e.g., goods produced by a factory are stored until transported in trucks). The *received* stage represents the phase in which things flowing from other machines *arrive* to be *accepted* inside the machine or sent back to outside (e.g., wrong address).

For simplicity sake, in the modeling examples in this paper we assume that things that arrive are always accepted, hence we always use the receive stage in these examples.

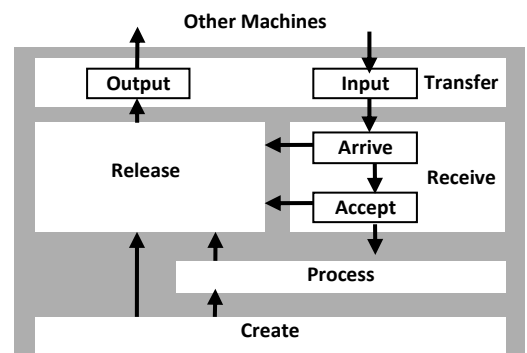

Fig. 1. The thinging machine model

The *create* stage in Fig. 1 denotes the appearance of a new thing in the machine (e.g., a generator's output after converting a form of energy into electricity). The process stage in Fig. 1 refers to changing the form of a thing without generating a new entity (e.g., transforming a decimal number to binary form). A stage in TM may include a storage area (represented by a cylinder) that accommodates things inside the stage. Fig. 2 shows a simplification of Fig. 1 where things flow from the output of one machine to the input of another machine.

A thing in TM modeling is whatever is created, processed, released, transferred, and received. Thus, a machine creates, processes, releases, transfers, and receives things. Hence, create, process, release, transfer, and receive are called *actions*.

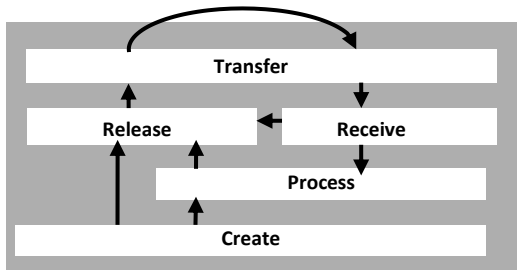

Fig. 2. Simplified TM model

The arrows in Fig. 1 denote the flow of things from one stage to another or within the machine. The TM model is the grand machine that results from smaller machines. To facilitate shifting among flows (e.g., processing electricity creates cold air), the TM model includes triggering denoted by dashed arrows.

## Example: Making a Bank Withdrawal

According to Elsayed et al. (2020), a UML sequence diagram is employed when dealing with the dynamic view of a system. A sequence diagram is one of the interaction diagrams that model the interactions between object instances and the sequential ordering of messages according to time. A vertical position indicates the sequence of the messages, wherever the first message is always shown at the top of the diagram. Elsayed et al. (2020) give an example of a sequence of messages.

### Static TM Model of the Withdrawal Process

Fig. 3 shows the corresponding TM model. First the user inserts his or her card (yellow circle; 1) and the card flows to the ATM (2) where it is received (3) and processed (4). This triggers (5) the creation of a message (6) to request the PIN input (7). Hence, the user inputs the PIN (8) that moves to the ATM (9) where it is processed (10) and sent to the bank (11).

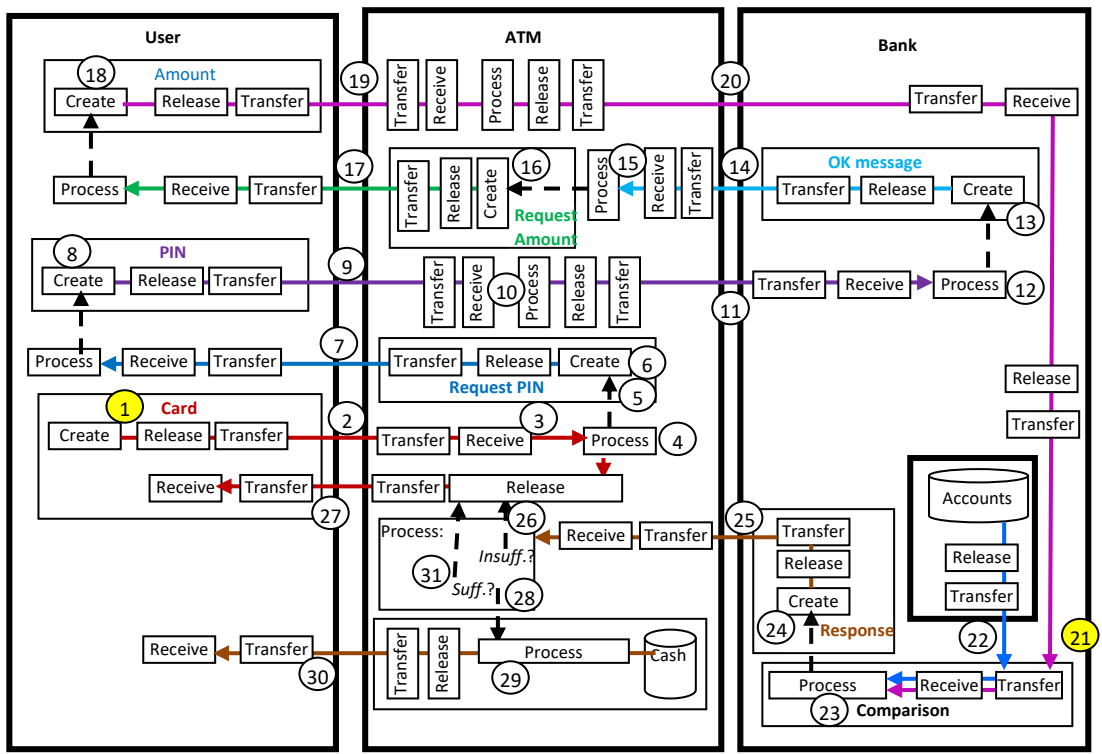

Fig. 3. The static TM model corresponding to the sequence diagram

■ ■

The bank processes the PIN (12) to generate an OK message (13). The OK message flows to the ATM (14) to be processed (15) and triggers the creation of a request to specify the amount (16) that goes to the user (17). The user inputs the amount (18) that flows to the ATM (19) and the ATM forwards the amount to the bank (20).

In the bank, the received amount (yellow circle, bottom right; 21) and the balance, retrieved from the database (22), are compared to generate the appropriate response (24). The response moves to the ATM (25) where it is processed:

- If the response reports insufficient funds (26), then the card is released back to the user (27).
- If sufficient funds (28) exist, cash is retrieved (29) and sent to the user (30), and the card is released to the user (31).

We claim that a shortcoming of the UML sequence diagram is that it models the behavior of the system based on the so-called events that are ordered vertically. We claim that this vertical order reflects a logical order and not a time-based order. In TM, an event is explicitly specified in fixed time. For example, the event that *the user has inserted a card that is received by the ATM* is modeled as shown in Fig. 4. The region in Fig. 4 denotes where the event occurs. For simplicity sake, we will represent an event by its region.

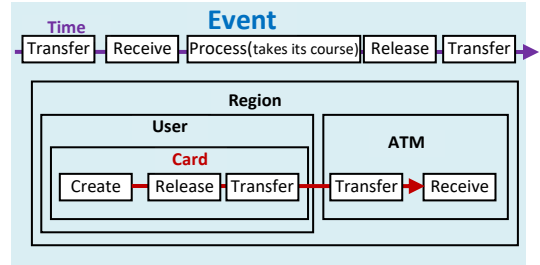

Fig. 4. The event *The User inserts a card that is received by the ATM at a specific time*

### The TM Events Model

To develop the system behavior, the static model (Fig. 3) is decomposed to identify event regions. Such decomposition depends on many factors because events may contain less general events. However, five generic events correspond to the five generic actions (i.e., create, process, etc.). Fig. 5 shows a sample decomposition of the ATM example as follows.

Event 1 ($E_1$): The user inserts his or her card that is received by the ATM.

Event 2 ($E_2$): The ATM processes the card and generates a request for the PIN.

Event 3 ($E_3$): The user inputs the PIN that is received and processed by the ATM.

Event 4 ($E_4$): The ATM sends the PIN to the bank to be processed.

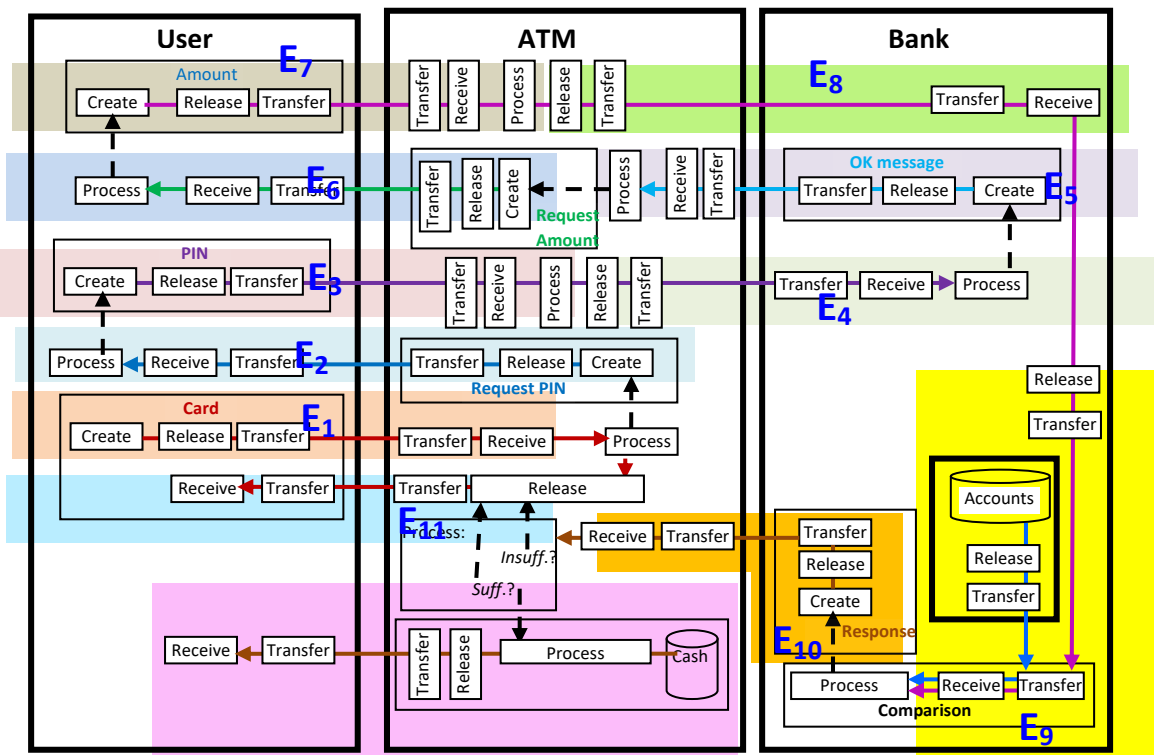

Fig. 5. The events in the TM model that correspond to the sequence diagram

∎∎

Event 5 (E$_5$): The bank sends OK message to the ATM.

Event 6 (E$_6$): The ATM requests the amount from the user.

Event 7 (E$_7$): The user inputs the amount that is received by the ATM.

Event 8 (E$_8$): The ATM sends the amount to the bank.

Event 9 (E$_9$): The bank compares the requested amount with the relevant account.

Event 10 (E$_{10}$): A bank response is created and sent to the ATM.

Event 11 (E$_{11}$): The ATM processes the response that reports the funds are insufficient, thus the card is returned to the user.

Event 12 (E$_{12}$): The ATM processes the response that reports funds are sufficient, thus cash is released to the user.

## Modeling the Little Man Computer

In LMC, we see, inside the CPU, a little man who runs around organizing 100 numbered mailboxes, a calculator, a two-digit counter, an in tray and an out tray, as shown in Fig. 6. We will not go through a complete description (e.g., address size or memory location size) of the LMC because it is clear for computer engineers and scientists. Bordewich (1965) described the little man's activities as follows.

1. The little man starts by looking at the counter for number, which is a mailbox number.

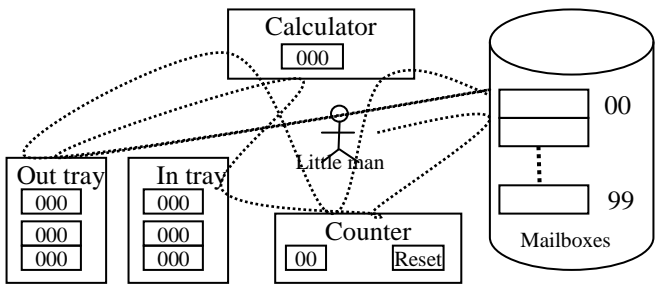

Fig. 6. Little man computer (Redrawn with modifications from Bordewich, 1965)

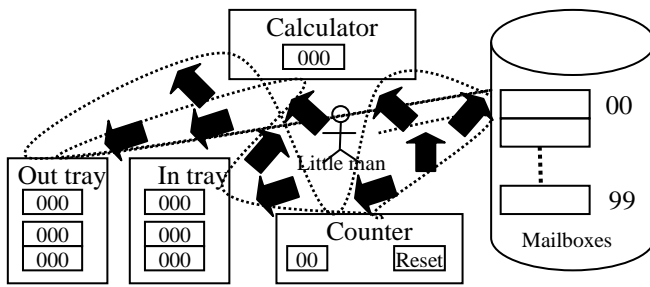

Fig. 7. Little man activities will be modeled

2. He increments the counter, so that next time he comes it will be one larger, and so on.

Instead, we will model the little man's activities as illustrated in Fig. 7.

### Static Model of LMC

Fig. 8 shows the static TM model of the LMC. According to Bordewich (1965), the LCM starts by looking at the counter for numbers (addresses):

- In Fig. 8, this corresponds to retrieving the value of the program counter (circle 1) shown in the form of drum at the lower left corner of the figure. The program counter is incremented (2) so that next time it will be one unit larger. Because this paper is read by computer science teachers, and to relate the material to actual computer terminology, we will mix the wording of LMC with technical words.

- The address in the program counter flows to the mail system (3) and the mailbox is processed (4) to fetch (5) the content (instruction) inside that location with that address.

- Note how the mail system is structured, the mailboxes comprise a whole cabinet (memory) that is then processed (red rectangles), which is different from a single memory content location that is released and transferred. Fig. 8 says that given an address, the memory is processed to extract the content of the box (location) with the given address.

- The fetched instruction is processed (6) to extract the opcode (7) and address inside the instruction. We assume here that the meaning of such words as *memory*, *instruction*, *opcode*, and *address* have already been explained to the student when explaining the figure.

- The opcode is processed (9) as follows.
  (a) If opcode = 0, then stop.
  (b) If opcode = 1 (add), then the address (10) is sent to the memory system where it is processed (4) to fetch the value that is stored in the address (11; green circle). This value flows to the calculator system (12) where it is added to the current value of the calculator to produce the new value of the calculator (13).
  (c) If opcode = 2 (circle 14) then as in (b) (circles 10, 4). The retrieved data (15) flows to the calculator system (16) where the value is subtracted from the calculator value (17). The result is examined (18), and if it is positive, then the result is stored in the calculator (19); otherwise, the negative flag is set (20) in addition to storing the result: Note that opcode 3 (store) will be explained later because of its position in the diagram.

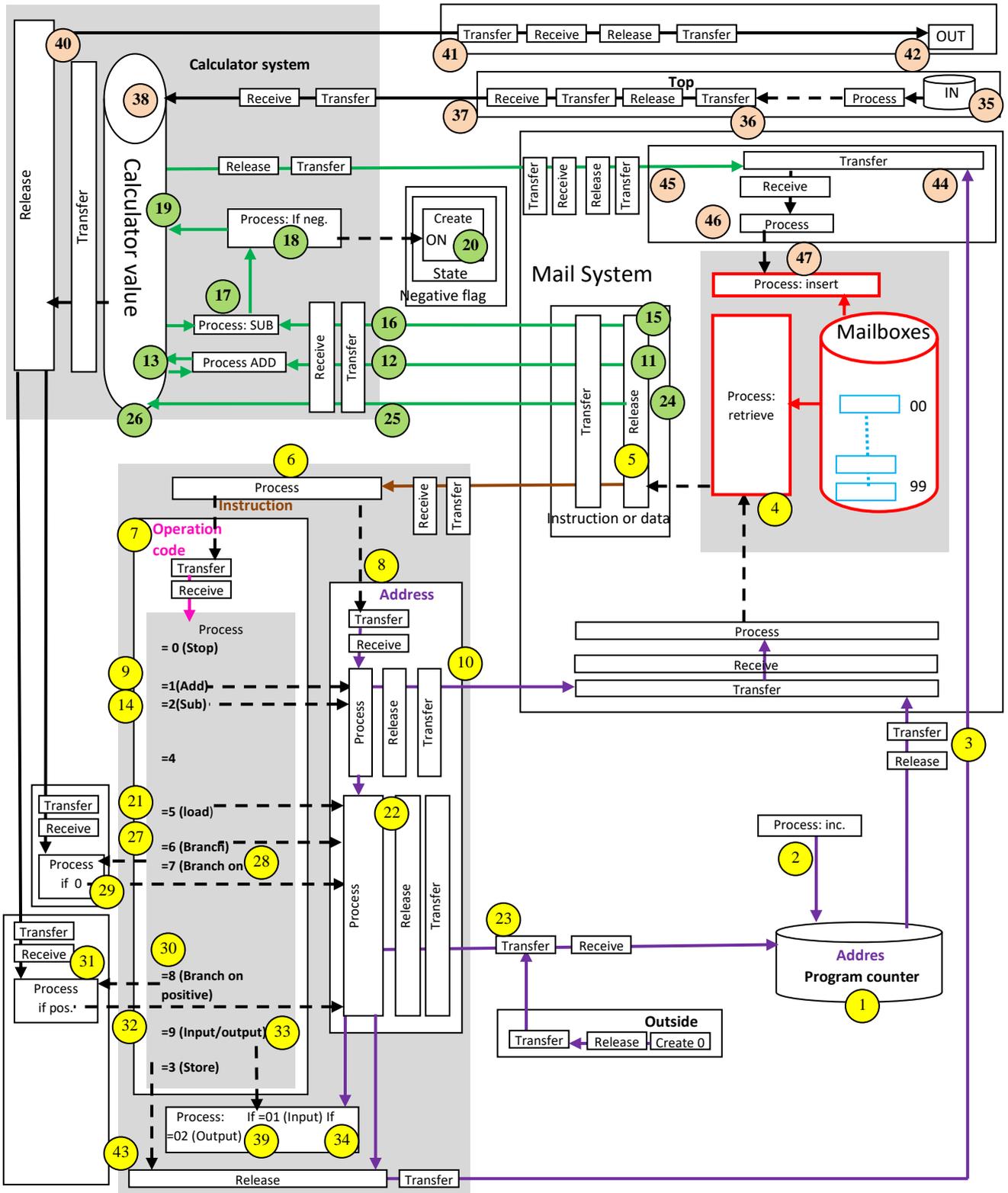

Fig. 8. The static TM model of the LMC

■■

To execute load, the corresponding address (22) is sent (23) to the program counter (1). The address then flows to the memory system (3) where the corresponding value is retrieved (4 and 24). Then this value becomes the new value of the calculator (25 and 26).

(e) If the opcode is 7 (e.g., branch; circle 27), then the address (22) flows (23) to the program counter (1). The address in the program counter is sent to the memory system (3) where another instruction is retrieved and processed (4, 5, and 6).

(f) If the opcode is 7 (branch on 0; circle 28), then the value of the calculator is processed (circle 28), and if it is zero, then the address (22) flows (29) to the program counter (1). The address in the program counter is sent to the memory system (3) where another instruction is retrieved and processed (4, 5, and 6).

(g) If the opcode is 8 (branch on positive; circle 30), then the value of the calculator is processed (circle 31), and if it is positive (32), then the address (22) flows (23) to the program counter (1). The address in the program counter is sent to the memory system (3) where another instruction is retrieved and processed (4, 5, and 6).

(h) If the opcode is 9 (input/output; 34), then the address field is examined.
  - If it is 01 (input; 34), then the input (35 [orange circle at the upper right corner]) is processed to extract its top data element (36) that flows (37) to the calculator (38).
  - If the address is 02 (39 [bottom left corner]), then the content of the calculator is released (40) to flow (41) to the output (42).

(i) If the opcode is 5 (store), then the address of the instruction is released (43 [lower left corner of the diagram]) and flows to where it is received (44 [upper right corner of the diagram]) along with the value of the calculator (45). The address and the calculator value are processed to trigger (46) the storing of the value in the memory according to the address (47).

*Events Model of LMC*

Accordingly, an events model is developed as follows (see Fig. 9).

Event 1 ($E_1$): The value zero is input from the outside.

Event 2 ($E_2$): The program counter is initialized to a new value.

Event 3 ($E_3$): The program counter is incremented.

Event 4 ($E_4$): The program counter value flows to the memory system.

Event 5 ($E_5$): The content of memory location (instruction) that correspond to the program counter is retrieved and sent to be processed.

Event 6 ($E_6$): The instruction processing produces the opcode and the address.

Event 7 ($E_7$): The opcode is processed and found to be 0.

Event 8 ($E_8$): The opcode is processed and found to be 1 (add).

Event 9 ($E_9$): The opcode is processed and found to be 2 (subtract).

Event 10 ($E_{10}$): The opcode is processed and found to be 3 (store).

Event 11 ($E_{11}$): The opcode is processed and found to be 4.

Event 12 ($E_{12}$): The opcode is processed and found to be 5 (load).

Event 13 ($E_{13}$): The opcode is processed and found to be 6 (branch)

Event 14 ($E_{14}$): The opcode is processed and found to be 7 (branch on 0).

Event 15 ($E_{15}$): The opcode is processed and found to be 8 (branch on positive).

Event 16 ($E_{16}$): The opcode is processed and found to be 9 (input/output).

Event 17 ($E_{17}$): The address is sent to the mail system.

Event 18 ($E_{18}$): The value of the mailbox location is retrieved and sent to the calculator where it is added to the calculator value.

Event 19 ($E_{19}$): The value of the mailbox location is retrieved and sent to the calculator where it is subtracted from the calculator value.

Event 20 ($E_{20}$): The result of subtraction is positive; hence, the value is stored in the calculator.

Event 21 ($E_{21}$): The result of subtraction is negative; hence, the negative flag is set ON and the value is stored in the calculator.

Event 22 ($E_{22}$): The value of the calculator is sent to the mailbox system.

Event 23 ($E_{23}$): The address is sent to the mail system.

Event 24 ($E_{24}$): The data incoming to the mailbox system ($E_{22}$) are stored in the memory according to the given address ($E_{23}$).

Event 25 ($E_{25}$): The address is sent to the memory system and the value of the location is loaded in the calculator.

Event 26 ($E_{26}$): The address is sent to the program counter.

Event 27 ($E_{27}$): The calculator value is processed and found to be 0.

Event 28 ($E_{28}$): The calculator value is processed and found to be positive.

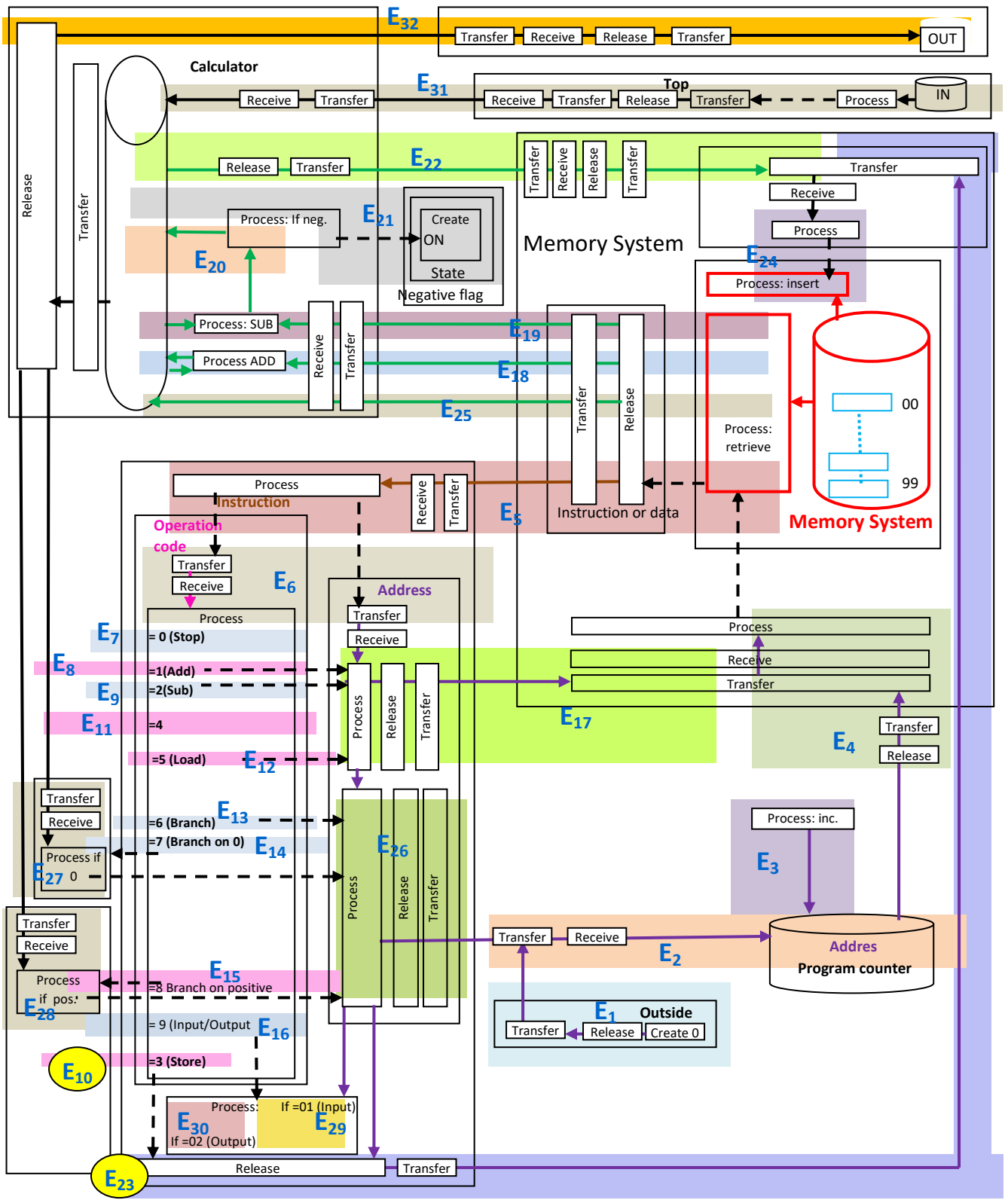

Fig. 9. The events of the TM model of the LMC

■■

Event 29 (E$_{29}$): The address is 01 (input).

Event 30 (E$_{30}$): The address is 02 (output).

Event 31 (E$_{31}$): Move the top of the input tray to the calculator.

Event 32 (E$_{32}$): Move the value of the calculator to the output tray.

### Events Model of LMC

Fig. 10 shows the TM behavioral model of the LMC. Consider the following program (from Bordewich, 1965):

```
IN
STO A
IN
ADD A
OUT
HLT
A DAT
```

Fig. 11 shows the behavior of such a program. The merit of such TM models can be recognized in the education field and program analysis field.

## Conclusion

In this study, we aimed mainly at developing a diagrammatic static/dynamic modeling of processes in the computer organization and architecture field.

We claim that the modeling of LMC gives initial indication of potentially wide application of TM in some areas of computer organization and architecture in such aspects as education, documentation, and analysis of processes. Such fruitful aspects can be confirmed with further similar research in modeling classical computer organization and architecture units. Other potential possible uses exist in processes that are more complex such as high-level design, which can be included in future studies.

One criticism directed at the TM model is that the diagrams are complex. Such an observation is superficial because it criticizes completeness. Imagine if electronic circuitry or airplane schemata were rejected for being complicated. Nevertheless, simplification can be applied to any level based on the complete diagram of the system and not before that. For example, the static TM model of LMC (Fig. 8) can be simplified greatly by erasing the release, transfer, and receive stages under the assumption that the direction of the arrows is sufficient to indicate the flow of thing (see Fig. 12). Further simplifications are possible for different purposes.

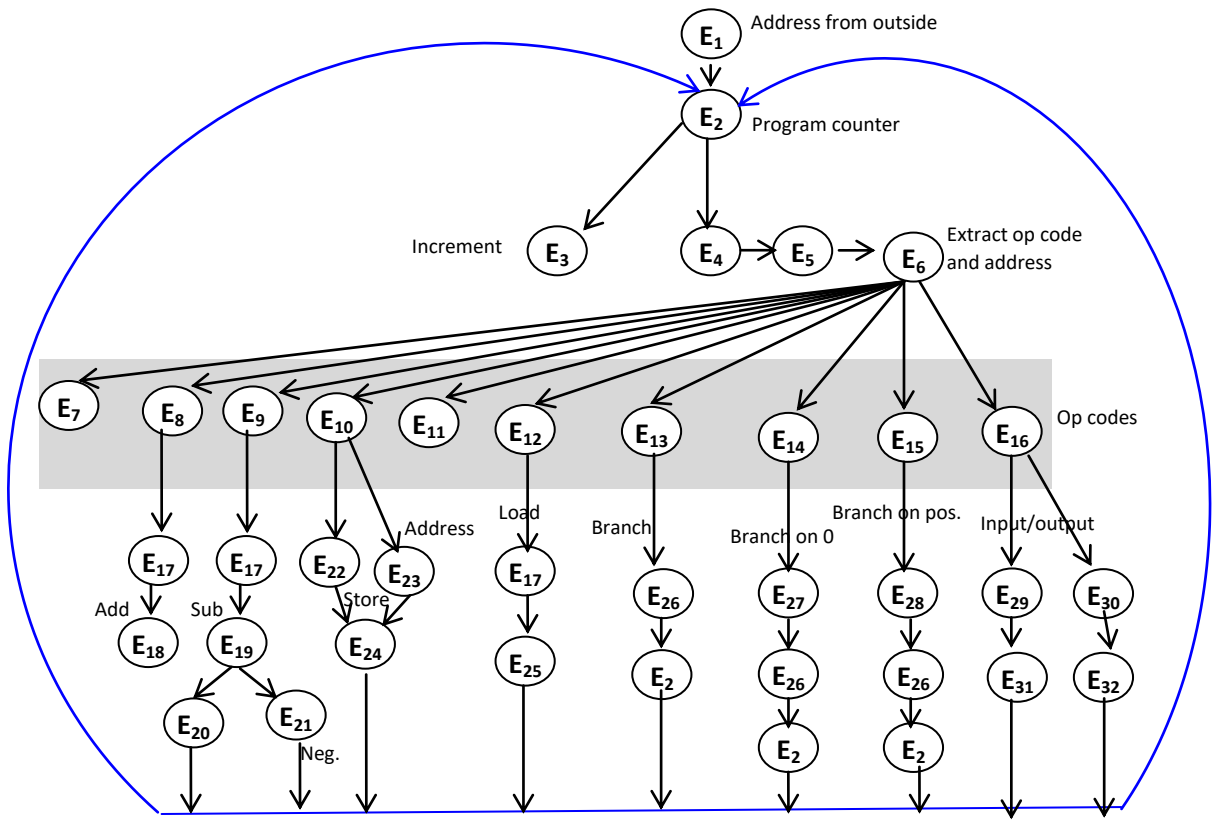

Fig. 10. The behavioral TM model of the LCM

■■

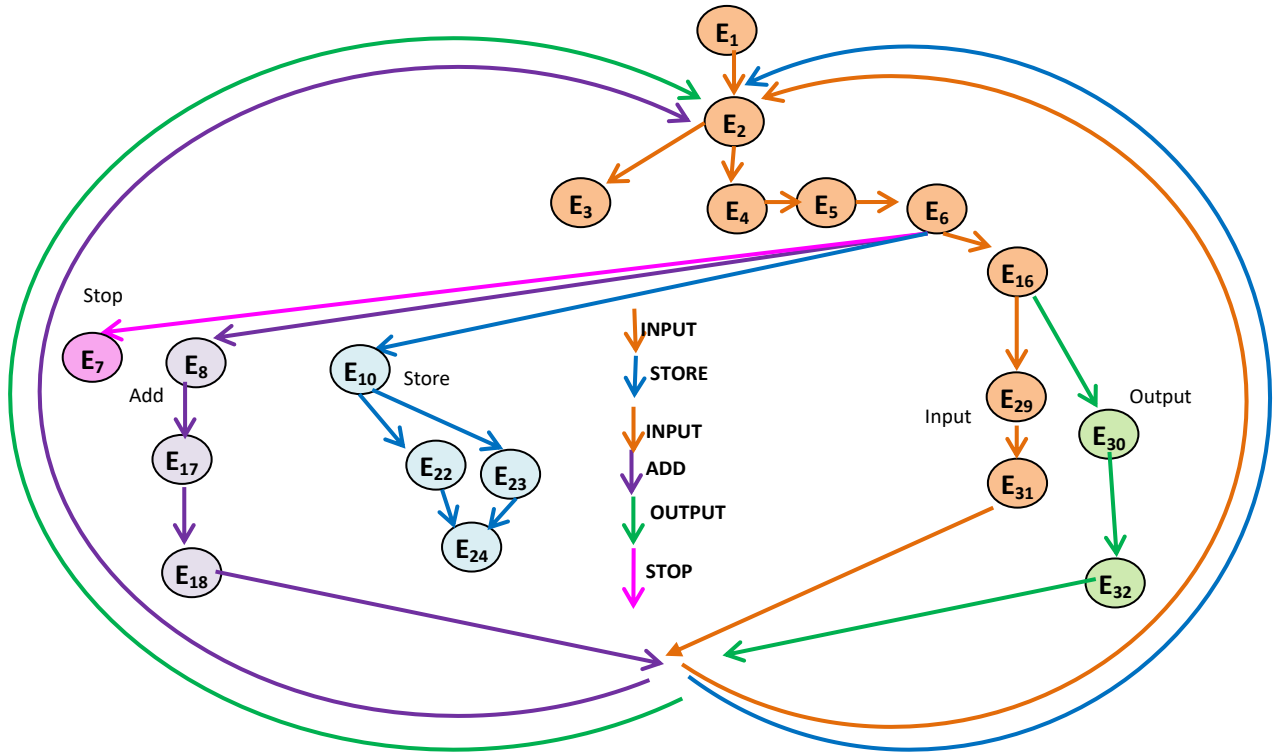

Fig. 11. The events in the sample program

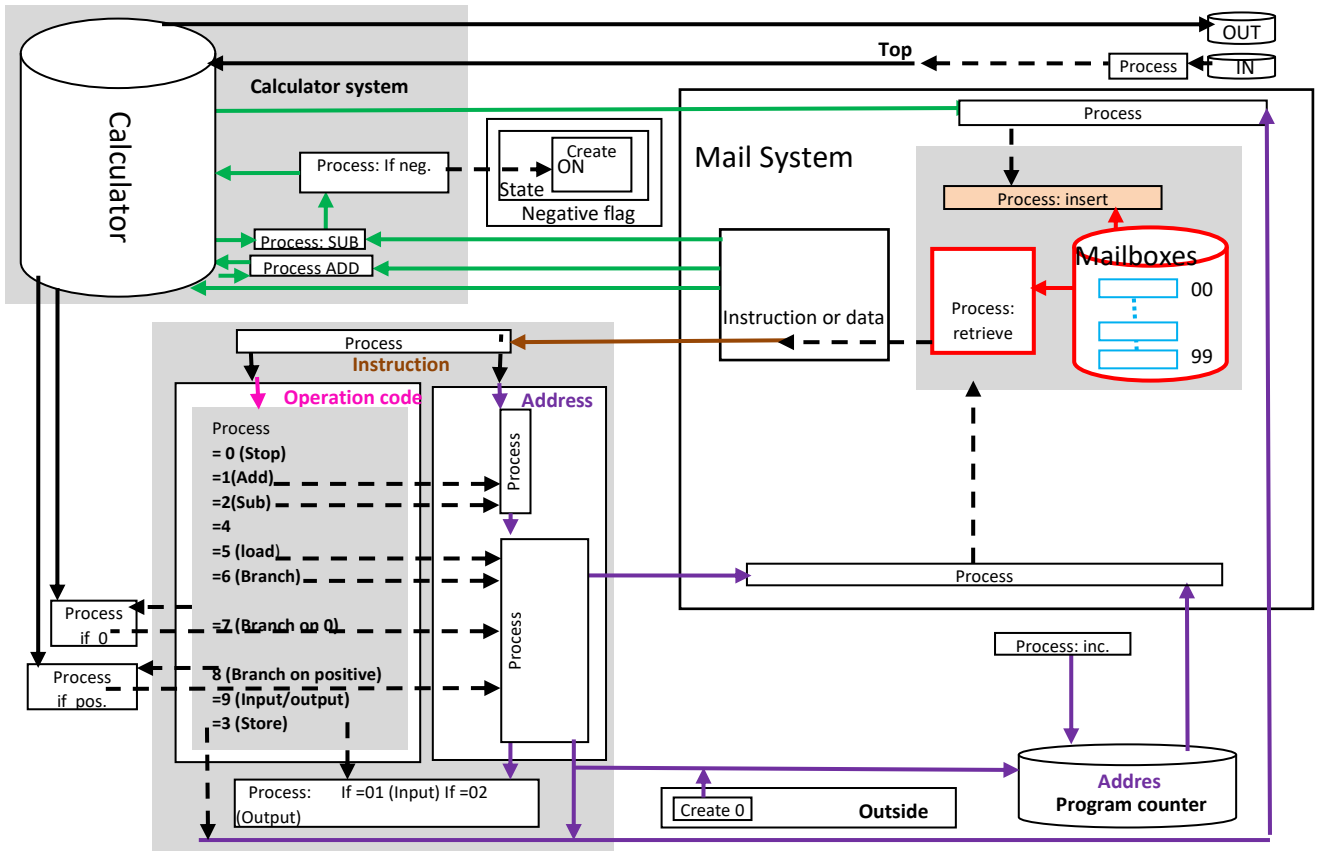

Fig. 12. Simplification of the static TM model of the LMC by removing the release, transfer, and receive stages

## Ethics

This article is original and contains unpublished material. No ethical issues were involved.